\documentclass[conference]{IEEEtran}
\IEEEoverridecommandlockouts
\usepackage{mathtools}
\usepackage{mathenv}
\usepackage{nccmath}

\usepackage{amsmath}
\usepackage{cuted}
\usepackage{cite}
\usepackage{amsmath,amssymb,amsfonts}
\usepackage{algorithmic}
\usepackage{algorithm}
\usepackage{graphicx}
\usepackage{textcomp}
\usepackage{xcolor}
\usepackage{subfigure}

\usepackage{booktabs,threeparttable}

\newcommand\ddfrac[2]{\frac{\displaystyle #1}{\displaystyle #2}}
\DeclareMathOperator*{\argmax}{arg\,max}

\def\BibTeX{{\rm B\kern-.05em{\sc i\kern-.025em b}\kern-.08em
    T\kern-.1667em\lower.7ex\hbox{E}\kern-.125emX}}
\begin{document}

\title{Scalar Quantizer Design for Two-Way Channels\\
\thanks{This work was supported in part by NSERC of Canada.}
}

\author{\IEEEauthorblockN{Saeed Rezazadeh, Fady Alajaji, and Wai-Yip Chan}
	\IEEEauthorblockA{Queen’s University, Kingston, ON, K7L 3N6 \\
		Email: s.rezazadeh@queensu.ca, fady@mast.queensu.ca, chan@queensu.ca}
}

\maketitle

\begin{abstract}
The problem of lossy transmission of correlated sources over memoryless two-way channels (TWCs) is considered. The objective is to develop a robust low delay and low complexity source-channel coding scheme without using error correction. A simple full-duplex channel optimized scalar quantization (COSQ) scheme that implicitly mitigates TWC interference is designed. Numerical results for sending Gaussian bivariate sources over binary additive-noise TWCs with either additive or multiplicative user interference show that, in terms of signal-to-distortion ratio performance, the proposed full-duplex COSQ scheme compares favourably with half-duplex COSQ. 
\end{abstract}


\section{Introduction}\label{sec:introduction}
For single-user communication systems, Shannon's separation source-channel coding theorem \cite{b1,b2} states that performing data compression and error control separately is optimal for reliable communication.
Although this result simplifies communication system design, the requirement of asymptotically large codeword lengths introduces serious decoding latency making the design inapplicable for systems with complexity and delay constraints such as wireless links. Moreover, the separation principle may not hold for multi-user communication even under unlimited resources \cite{b16}. Over the last decades, various efficient joint source-channel coding (JSCC) schemes were developed (see \cite[Section 4.6]{b22} and the references therein) to address the above problems. The aim of this paper is to construct a robust, low-delay, and low-complexity JSCC scheme for two-user two-way channels (TWCs) \cite{b3}.

The TWC is a fundamental two-user model that allows both users to transmit information in a full-duplex manner, thus improving the spectral efficiency with respective to one-way systems (i.e., half-duplex transmission). Moreover, the TWC setup inherently enables users to cooperate by adapting channel inputs to previously received signals, potentially resulting in lower end-to-end errors or distortions. Information-theoretical studies for TWCs were made from different perspectives. In~\cite{b12}, Kaspi investigated the lossy source coding problem over a noiseless TWC, in which the information is transferred only through a one-way link, establishing a rate-distortion region. Maor et al. later adopted the interactive protocol proposed in \cite{b12} and developed a lossy transmission scheme for noisy TWCs\cite{b13}. Moreover, scalar quantize and forward designs for the two-way relay channel were studied in \cite{b20} for systems where the two terminals exchange source data with the help of a relay when there is no direct link between the terminals. Achievability and converse results for the lossy transmission of correlated sources over TWCs were derived in \cite{b4}. In the same paper, the optimality of scalar coding for TWCs with discrete modulo additive noise as well as additive white Gaussian noise was investigated. Considering that a TWC can be described as two state-dependent one-way channels, the authors in \cite{b5} extended the one-way hybrid digital/analog coding scheme of \cite{b6} to TWCs.

For one-way channels, various JSCC techniques have been studied. For instance, JSCC schemes were examined for image coding \cite{b8,b9} as well as speech coding \cite{b10, b11} applications. Also in \cite{b17}, the problem of JSCC is addressed when variable length codes are used over discrete memoryless channels. Channel optimized scalar quantization (COSQ) is a well-known robust lossy JSCC scheme with low complexity and low delay \cite{b7} in which the source encoder is a zero-memory quantizer. In \cite{b18}, joint  source-channel scalar quantizers  were studied in conjunction with turbo-codes; the proposed design showed superior performance compared to conventional COSQ schemes \cite{b7} by using a soft reconstruction of samples at the decoders. In the multi-user setup, the authors in \cite{b19} used multi-resolution quantization and layered JSCC to serve simultaneously several users over the binary erasure broadcast channel. Also, in \cite{b24}, the proposed single-user JSCC scheme of \cite{b25} was adopted for the transmission of two correlated sources over an orthogonal multiple-access channel. In this paper, we propose a COSQ design for the effective transmission of generally correlated sources over discrete memoryless two-way channels.    

To the best of our knowledge, the problem of COSQ design over TWCs has not been studied. In this work, we extend the optimality conditions of \cite{b7} to TWCs so that our proposed scheme judiciously mitigates the self-interference caused by either users and exploits the statistical dependency between the users as receivers' side information. As a first step, we consider a restricted TWC that does not adapt the channel inputs to prior received outputs.

The rest of this paper is organized as follows. In Section~\ref{sec:notation_prob_stat}, the proposed design is introduced. The necessary conditions for system optimality are derived in Section~\ref{sec:necessary_conditions}. Numerical results and comparisons with proposed baseline systems are presented in Section~\ref{sec:results_com}. Finally, Section~\ref{sec:conclusion_future} concludes the paper.    
\section{COSQ design for TWCs}\label{sec:notation_prob_stat}

\begin{figure}[h]
	\centering
	\includegraphics[scale=0.65]{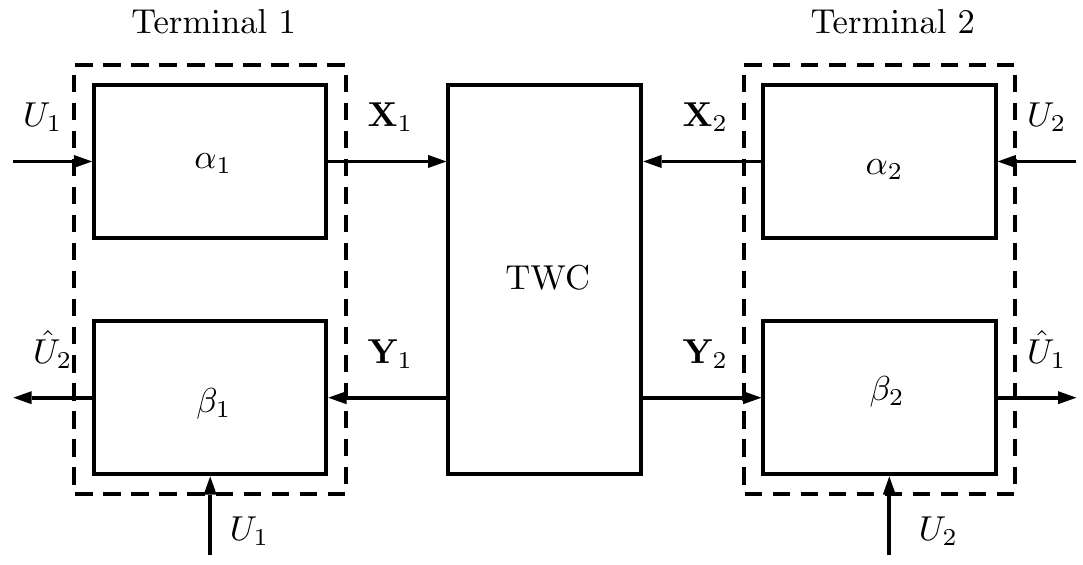}
	\caption{The block diagram of the COSQ TWC system.}
	\label{fig:block_diagram}
\end{figure}

In a two-way communication system, two users wish to simultaneously exchange source data over a discrete memoryless TWC. The block diagram of the proposed system is depicted in Fig.~\ref{fig:block_diagram}. At terminal $j$, the input source to the COSQ encoder is a real-valued memoryless process $\{U_{j,n}\}_{n = 1} ^ {\infty}$ for $j = 1 , 2$. We assume that at each time instant $n$, the source samples $U_{1,n}$ and $U_{2,n}$ are correlated in general. For the sake of convenience, as we are dealing with time-memoryless sources, we drop the time index $n$ and write $U_{j,n}$ as $U_j, j = 1 , 2$. The corresponding COSQ encoder at terminal $j$ is a mapping $\alpha_j$ that takes a source realization $u_j \in \mathbb{R}$ and outputs an index $N$-tuple $\textbf{x}_j = (x_{j1} , \ldots , x_{jN})\in {\cal X}^N$ where $\mathcal{X}$ is the channel input alphabet such that 
\begin{equation}
\alpha_j(u_j) = \textbf{x}_j, \vspace{0.1 in}\text{ if } u_j \in S_{\textbf{x}_j}
\label{eq:encoder}
\end{equation}
where ${\cal P}_j = \{S_{\textbf{x}_j},\textbf{x}_j \in {\cal X}^N\}$ is a partition of $\mathbb{R}$. The $N$-tuples $\textbf{x}_j \in {\cal X}^N$ are then 
transmitted via $N$ uses of a discrete memoryless TWC (used without adaptation) with the transition distribution $P_{{Y}_1,{Y}_2 | {X}_1 , {X}_2}$. For such a channel we have that $P_{\textbf{Y}_1,\textbf{Y}_2 | \textbf{X}_1 , \textbf{X}_2} (\textbf{y}_1 , \textbf{y}_2 | \textbf{x}_1 , \textbf{x}_2) = \prod_{i = 1}^{N}P_{Y_{1},Y_{2} | X_{1},X_{2}}(y_{1i},y_{2i} | x_{1i},x_{2i})$ where for $j = 1 , 2$, $\textbf{y}_j=(y_{j1} , \ldots , y_{jN})\in {\cal Y}^N$ is the received sequence at terminal $j$ and $\mathcal{Y}$ is the channel output alphabet. Finally, the corresponding decoder at terminal $j$ is a mapping $\beta_j : {\cal Y}^N\times \mathbb{R} \rightarrow \mathbb{R}$ that maps the received $N$-tuple $\textbf{y}_j$ to the output levels of a quantizer codebook using the statistical dependency between sources at the two terminals as side information: 
\begin{equation}
\beta_j (\textbf{y}_j , u_j) = c_{\textbf{y}_j , u_j}, \hspace{0.2 in}c_{\textbf{y}_j,u_j} \in \mathbb{R}, \hspace{0.08 in} \textbf{y}_j \in {\cal Y}^N, \hspace{0.08 in} u_j \in \mathbb{R}.
\label{eq:decoder}
\end{equation}

The proposed COSQ over the discrete memoryless TWC, aims to select the codebooks ${\cal C}_j = \{c_{\textbf{y}_j,u_j} , \textbf{y}_j \in {\cal Y}^N,  u_j \in \mathbb{R}\}$ and the partition sets ${\cal P}_j = \{S_{\textbf{x}_j},\textbf{x}_j \in {\cal X}^N\}$ to minimize the overall average mean-square error (MSE) given by: 
\begin{equation}
D = E((U_1 - \hat{U}_1) ^ 2 + (U_2 - \hat{U}_2) ^ 2)
\label{eq:MSE}
\end{equation}
where $\hat{U}_k$ is the reconstruction of the source $U_k, k = 1 , 2$. Hence, the overall MSE is described as:    
\begin{multline}
D =
\sum_{\textbf{x}_1}\sum_{\textbf{x}_2}\int_{S_{\textbf{x}_1}}\int_{S_{\textbf{x}_2}}f_{U_1 , U_2}(u_1 , u_2)\\ \times\sum_{\textbf{y}_1}\sum_{\textbf{y}_2}P_{\textbf{Y}_1 , \textbf{Y}_2 | \textbf{X}_1 , \textbf{X}_2}(\textbf{y}_1 , \textbf{y}_2 | \textbf{x}_1 , \textbf{x}_2)\nonumber\\\times\{(u_1 ~- ~ c_{\textbf{y}_2,u_2})^2 + (u_2 ~-~ c_{\textbf{y}_1 , u_1})^2\}\mathrm{d}u_1\mathrm{d}u_2
\label{eq:distoriton}
\end{multline}
where $f_{U_1 , U_2}$ is the joint probability density function of the two sources. The performance of this system is usually measured via the average distortion $D$ and the encoding rate given by: 
\begin{equation}
r = \log_2~M \hspace{0.1 in} \text{bits/source symbol}
\end{equation}
where $M = |{\cal X}|^ N$ is the total number of quantization indices.
\section{Necessary Conditions of Optimality}\label{sec:necessary_conditions}
For a given source, channel, and fixed rate $r$, we wish to find the optimal functions $\alpha_j$ and $\beta_j$ for $j = 1 , 2$ to minimize the overall MSE given by~\eqref{eq:distoriton}. The necessary conditions for optimal encoders and decoders are next determined. The optimal encoder for each user is derived for given decoders and the encoder of the other user. Similarly, the optimal decoder for each user is obtained given fixed encoders. Assuming known and fixed decoders and the encoder of user two, we seek the best encoding function for user one. Rewriting the overall MSE in~\eqref{eq:MSE} as: 
\begin{multline}
D = \int_{-\infty}^{\infty}f_{U_1}(u_1)\\\times E\{(U_1 - \hat{U}_1) ^ 2 + (U_2 - \hat{U}_2) ^ 2 | U_1 = u_1\}\mathrm{d}u_1,
\end{multline} 
we note by~\eqref{eq:encoder} that 
\begin{multline}
E\{(U_1 - \hat{U}_1) ^ 2 + (U_2 - \hat{U}_2) ^ 2 | U_1 = u_1\} \\= E\{(U_1 - \hat{U}_1) ^ 2 + (U_2 - \hat{U}_2) ^ 2 | \textbf{X}_1 = \textbf{x}_1 , U_1 = u_1\}.
\end{multline}
To minimize $D$, it is sufficient to find a mapping $\alpha_1$ that minimizes the modified distortion measure defined as: $d_1(u_1 , \textbf{x}_1) \triangleq E\{(U_1 - \hat{U}_1) ^ 2 + (U_2 - \hat{U}_2) ^ 2 | \textbf{X}_1 = \textbf{x}_1 , U_1 = u_1\}$. Therefore, given fixed codebooks ${\cal C}_1$, ${\cal C}_2$, and the partition set, ${\cal P}_2$, for user two, $D$ is minimized provided that the partition ${\cal P}_1$ satisfies:
\begin{subequations}
\begin{multline}
S_{\textbf{x}_1} = \{u_1 : d_1(u_1 , \textbf{x}_1) \\\le d_1(u_1 , \hat{\textbf{x}}_1), \hspace{0.1 in} \hat{\textbf{x}}_1 \in {\cal X}^N\}, \hspace{0.1 in} \textbf{x}_1 \in {\cal X}^N
\label{eq:encoder_1}
\end{multline}  
where 
\begin{multline}
d_1(u_1 , \textbf{x}_1) = \sum_{\textbf{x}_2}\sum_{\textbf{y}_1}\sum_{\textbf{y}_2}P_{\textbf{Y}_1 , \textbf{Y}_2 | \textbf{X}_1 , \textbf{X}_2}(\textbf{y}_1 , \textbf{y}_2 | \textbf{x}_1 , \textbf{x}_2)\\\times\int_{S_{\textbf{x}_2}}f_{U_2 | U_1}(u_2 | u_1)((u_1 - c_{\textbf{y}_2,u_2}) ^ 2 + (u_2 - c_{\textbf{y}_1 , u_1}) ^ 2)\mathrm{d}u_2.
\end{multline}

Similarly, given fixed codebooks ${\cal C}_1$, ${\cal C}_2$, and the partition set, ${\cal P}_1$, for user one, the average distortion $D$ is minimized if the partition ${\cal P}_2$ satisfies: 
\begin{multline}
S_{\textbf{x}_2} = \{u_2 : d_2(u_2 , \textbf{x}_2 ) \\\le d_2(u_2 , \hat{\textbf{x}}_2), \hspace{0.1 in} \hat{\textbf{x}}_2 \in {\cal X}^N\}, \hspace{0.1 in} \textbf{x}_2 \in {\cal X}^N
\label{eq:encoder_2}
\end{multline}
where 
\begin{multline}
d_2(u_2 , \textbf{x}_2) \triangleq\\
\begin{aligned}
   E\{(U_1 - \hat{U}_1) ^ 2 + (U_2 - \hat{U}_2) ^ 2 | \textbf{X}_2 = \textbf{x}_2 , U_2 = u_2\}\\=
\sum_{\textbf{x}_1}\sum_{\textbf{y}_1}\sum_{\textbf{y}_2}P_{\textbf{Y}_1 , \textbf{Y}_2 | \textbf{X}_1 , \textbf{X}_2}(\textbf{y}_1 , \textbf{y}_2 | \textbf{x}_1 , \textbf{x}_2)\\\times\int_{S_{\textbf{x}_1}}f_{U_1 | U_2}(u_1 | u_2)((u_1 - c_{\textbf{y}_2,u_2}) ^ 2 \\+ (u_2 - c_{\textbf{y}_1 , u_1}) ^ 2)\mathrm{d}u_1.
\end{aligned}
\end{multline}
\label{eq:encoders}
\end{subequations}

It can be shown that the optimal decoders for fixed and known encoders are the conditional expectation of the source, given the channel outputs and the locally observed source samples, i.e., 
\begin{subequations}
	\begin{align}
\begin{split}
\hat{u}_1 &= c^*_{\textbf{y}_2 , u_2} \\
&= E (U_1 | \textbf{Y}_2 = \textbf{y}_2 , U_2 = u_2)\\&=\medmath{\ddfrac{\sum_{\textbf{x}_1}P_{\textbf{Y}_2 | \textbf{X}_1 , \textbf{X}_2}(\textbf{y}_2 | \textbf{x}_1 , \alpha_2(u_2))\int_{S_{\textbf{x}_1}}u_1f_{U_1 | U_2}(u_1 | u_2)\mathrm{d}u_1}{\sum_{\textbf{x}_1}P_{\textbf{Y}_2 | \textbf{X}_1 , \textbf{X}_2}(\textbf{y}_2 | \textbf{x}_1 , \alpha_2(u_2))\int_{S_{\textbf{x}_1}}f_{U_1 | U_2}(u_1 | u_2)\mathrm{d}u_1}}
\end{split}
\label{eq:decoder_1}
\end{align}  
and 
\begin{align}
\begin{split}
\hat{u}_2 &= c^*_{\textbf{y}_1 , u_1} \\&=E (U_2 | \textbf{Y}_1 = \textbf{y}_1 , U_1 = u_1)\\&=\medmath{\ddfrac{\sum_{\textbf{x}_2}P_{\textbf{Y}_1 | \textbf{X}_1 , \textbf{X}_2}(\textbf{y}_1 | \alpha_1(u_1) , \textbf{x}_2)\int_{S_{\textbf{x}_2}}u_2f_{U_2 | U_1}(u_2 | u_1)\mathrm{d}u_2}{\sum_{\textbf{x}_2}P_{\textbf{Y}_1 | \textbf{X}_1 , \textbf{X}_2}(\textbf{y}_1 | \alpha_1(u_1) , \textbf{x}_2)\int_{S_{\textbf{x}_2}}f_{U_2 | U_1}(u_2| u_1)\mathrm{d}u_2}}.
\end{split}
\label{eq:decoder_2}
\end{align}
\label{eq:decoders}
\end{subequations}

Successive application of \eqref{eq:encoders} and \eqref{eq:decoders} forms a number of encoder-decoder pairs  resulting in a sequence of average distortion values. Given the structure of decoders in~\eqref{eq:decoders}, optimizing the decoders cannot increase the overall distortion. However, optimizing  the encoder at terminal $j$ for $j = 1 , 2$ requires a fixed encoder at the other terminal (see ~\eqref{eq:encoders}). This inherent entanglement in the structure of the two encoders may increase the distortion value from one iteration to the next during the optimization procedure. Although there in no guarantee to have monotonically decreasing distortion values as a function of iteration index, the distortion, for the cases we studied in this paper, did not increase and the algorithm converged. 
The proposed iterative COSQ algorithm is as follows: 
\begin{enumerate}
	\item Input the joint source pdf $f_{U_1 , U_2}$, the initial partition sets ${\cal P}_1^{(0)} = \{S^{(0)}_{\textbf{x}_1} , \textbf{x}_1 \in {\cal X}^N\}$, ${\cal P}_2^{(0)} = \{S^{(0)}_{\textbf{x}_2} , \textbf{x}_2 \in {\cal X}^N\}$, and a stopping threshold $T$.\\
	Set $m = 0$ and $D_0 = \infty$.
	\item \begin{itemize}
		\item Given the partitions ${\cal P}_1^{(m)}$ and ${\cal P}_2^{(m)}$, use~\eqref{eq:decoders} to get the optimal codebooks ${\cal C}_1^{(m)}$ and ${\cal C}_2^{(m)}$. 
		\item Given ${\cal C}_1^{(m)}$ and ${\cal C}_2^{(m)}$, use ~\eqref{eq:encoder_1} for a fixed ${\cal P}^{(m)}_2$ to find the optimal partition ${\cal P}^{(m + 1)}_1$. Similarly, given ${\cal C}_1^{(m)}$ and ${\cal C}_2^{(m)}$, use ~\eqref{eq:encoder_2} for an updated and fixed ${\cal P}^{(m + 1)}_1$ to find the optimal partition ${\cal P}^{(m + 1)}_2$.
	\end{itemize}
	\item Update the distortion $D_{m + 1}$ associated with the partitions ${\cal P}^{(m + 1)}_1$, ${\cal P}^{(m + 1)}_2$, and codebooks ${\cal C}_1^{(m)}$ and ${\cal C}_2^{(m)}$.
	\begin{itemize}
		\item If $\frac{D_{m} - D_{m + 1}}{D_{m}} \le T$, stop and output ${\cal C}_1^{(m)}$, ${\cal C}_2^{(m)}$, ${\cal P}^{(m + 1)}_1$, and ${\cal P}^{(m + 1)}_2$.
		\item  Otherwise, $m = m + 1$ and go to step (2).
	\end{itemize}
\end{enumerate}

We note that the proposed COSQ design implicitly optimizes the mapping from the quantization indices to the channel inputs as the encoders at each terminal directly map source symbols to channel inputs.

\section{Numerical Results}\label{sec:results_com}
In this section we present numerical results for two types of discrete memoryless TWCs: the binary-additive TWC (BA-TWC) with additive noise and the binary-multiplying TWC (BM-TWC) with additive noise where the latter is an extension of Blackwell's classical binary multiplying channel \cite{b3}. 
The outputs of the BA-TWC with additive noise at time $i \in \{1 , \ldots, N\}$ can be described as modulo-2 sum of its inputs and noise variables: 
\begin{equation}
\begin{cases}
Y_{1i} = X_{1i} \oplus X_{2i} \oplus Z_{1i},\\
Y_{2i} = X_{1i} \oplus X_{2i} \oplus Z_{2i},
\end{cases}
\label{eq:ANTWC}
\end{equation}  
where $\oplus$ denotes the addition modulo-2, $Y_{ji}\text{, } X_{ji} \text{, and } Z_{ji}$ are the channel outputs, inputs and noise variables at terminal $j$, respectively. The alphabets ${\cal X} = {\cal Y} = {\cal Z} = \{0 , 1\}$ are all binary. The noise variables $Z_1$ and $Z_2$, which are assumed to be independent of each other and of the channel inputs, have the following distributions: 
\begin{equation}
P_{Z_j} (z_j = 1) = \epsilon_j
\label{eq:noise}
\end{equation}
where $0 \le \epsilon_j < 1/2$ for $j = 1 , 2$.  
Similarly, the BM-TWC with additive noise, at time $i$, can be described as:
\begin{equation}
\begin{cases}
Y_{1i} = X_{1i} X_{2i} \oplus Z_{1i},\\
Y_{2i} = X_{1i} X_{2i} \oplus Z_{2i},
\end{cases}
\label{eq:BMC}
\end{equation}  
where the channel outputs, inputs, and noise processes have the same alphabets as described for the BA-TWC with additive noise. The noise variables for~\eqref{eq:BMC} are identical to those in~\eqref{eq:ANTWC} and~\eqref{eq:noise}. 

We have tested our proposed design for a bi-variate Gaussian source $(U_1 , U_2) $ with covariance matrix $$\Sigma = \begin{pmatrix} 
1 & \rho \\
\rho & 1 
\end{pmatrix}$$ 
where $\rho$ is the correlation coefficient of the two sources.

\subsection{Half-duplex COSQ System}
\begin{figure}[h]
	\centering
	\includegraphics[scale=0.65]{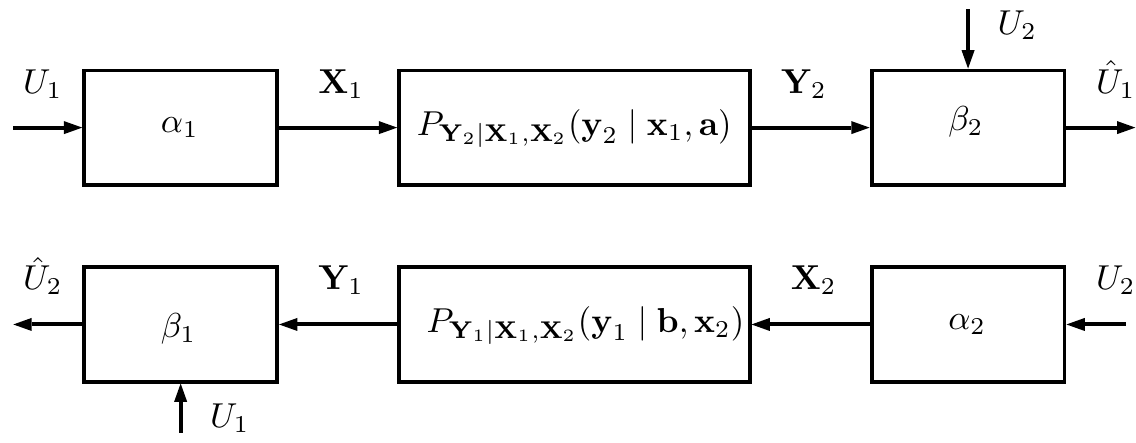}
	\caption{The block diagram of the half-duplex COSQ system where $\textbf{a} = (a^* , \ldots , a^*) \in {\cal X}^N$ and $\textbf{b} = (b^* , \ldots , b^*) \in {\cal X}^N$.}
	\label{fig:block_diagram_2}
\end{figure}

So far, we have proposed a COSQ design where the two users exchange sources simultaneously which we refer to as the full-duplex system. We also propose another design such that when one terminal is using the TWC, the other user transmits a constant symbol. We refer to the latter approach as the half-duplex system. The block diagram of the half-duplex design is depicted in Fig.~\ref{fig:block_diagram_2} where the marginal channel distributions are given by: $P_{\textbf{Y}_1 | \textbf{X}_1, \textbf{X}_2} (\textbf{y}_1 | \textbf{x}_1 , \textbf{x}_2) = \sum_{\textbf{y}_2}P_{\textbf{Y}_1 , \textbf{Y}_2 | \textbf{X}_1, \textbf{X}_2} (\textbf{y}_1, \textbf{y}_2 | \textbf{x}_1 , \textbf{x}_2)$ and $P_{\textbf{Y}_2 | \textbf{X}_1, \textbf{X}_2} (\textbf{y}_2 | \textbf{x}_1 , \textbf{x}_2) = \sum_{\textbf{y}_1}P_{\textbf{Y}_1 , \textbf{Y}_2 | \textbf{X}_1, \textbf{X}_2} (\textbf{y}_1, \textbf{y}_2 | \textbf{x}_1 , \textbf{x}_2)$. 

Considering a general TWC, the optimal symbol, in terms of maximal one-way information transfer, to be constantly applied by user two while user one is transmitting its source symbols is obtained by: 
\begin{equation}
a^* = \argmax_{{a \in {\cal X}}} \max_{P_{X_1 | X_2}(x_1 | a)} I(X_1 ; Y_2 | X_2 = a)
\end{equation}         
where $P_{X_1 | X_2}(x_1 | a)$ is the distribution of the channel inputs generated by user one given that the other user is transmitting $X_2 = a$, and $I(X_1;Y_2 | X_2 = a)$ is the conditional mutual information between $X_1$ and $Y_2$ given $X_2 = a$. As a result, the $N$-tuple that is transmitted by user two for $N$ consecutive uses of the TWC is given as: $\textbf{a} \triangleq (a^*, \ldots, a^*)$. Similarly, the optimal value that user one sends in the half-duplex mode is given by:   
\begin{equation}
b^* = \argmax_{{b \in {\cal X}}} \max_{P_{X_2 | X_1}(x_2 | b)} I(X_2 ; Y_1 | X_1 = b),
\end{equation}
and the optimal $N$-tuple for user one is: $\textbf{b} \triangleq (b^* , \ldots, b^*)$.
The proposed half-duplex design can be treated as two independent COSQ systems for two one-way channels. We adopt the iterative algorithm along with the optimality conditions proposed in \cite{b7}, to design the optimal quantizers for half-duplex schemes. 

Moreover, for the BA-TWC with additive noise, we choose the partition sets obtained from the half-duplex design as the initial partition sets for the full-duplex scheme with the same source and channel parameters. 

However, for the BM-TWC the ultimate partition sets from the half-duplex scheme are used to initialize the full-duplex design with twice the quantization rate. For every source sample, the $N / 2$-tuple partition indices of the half-duplex design is left-padded by all-one $N/2$-tuples for user one. The initial partition indices of the user two in the full-duplex design is obtained in the same way, however, by right-padding the partition indices of the half-duplex scheme with all-one $N/2$-tuples. This method of initialization provides each users symbols with protection against the other user's transmission. Afterwards, we gradually increase the channel noise parameters $\epsilon_1$ and $\epsilon_2$, according to the values in Tables \ref{table:SDR_ANTWC_p} and \ref{table:SDR_BMC_p}, while each time the previously found partitions are set as initializations.    
	
\subsection{Results and Discussion}
Tables~\ref{table:SDR_ANTWC_p} and \ref{table:SDR_BMC_p} show the COSQ performance results for full-duplex designs compared to the corresponding half-duplex schemes for additive-noise TWCs with additive and multiplicative user interference, respectively. The performance results are in terms of signal-to-distortion ratio (SDR) defined as: $\text{SDR} = 10\times\log_{10}~\frac{2}{D}$ (in dB), as both sources have unit variances. 

In Table I, we also include the SDR optimum performance theoretically achievable (OPTA) using the complete JSCC theorem in \cite[Theorem 3]{b5} (by calculating the sum of the users distortions) which readily applies for the case of correlated Gaussian sources sent over the BA-TWC with additive noise.  However in Table II, since there is no complete JSCC theorem for the BM-TWC (as the exact determination of its capacity region is still an open problem even in the absence of additive noise), we include an upper bound on the system’s OPTA using the converse result in \cite[Lemma 2]{b4}. The exact/upper bound OPTA values are meant to show the best performance potentially realizable if one were to employ powerful source-channel codes with unlimited delay and complexity. Naturally, since our low-delay COSQ schemes are scalar, their performance are considerably below the OPTA bound; this gap can however be reduced with the use of high-dimensional channel optimized vector quantizers in conjunction with capacity-achieving channel codes.  

Setting the quantization rate in the full-duplex system as twice the quantization rate in the half-duplex system makes the number of source symbols transmitted per the total number of TWC uses equal in both full-duplex and half-duplex systems. In this set-up, the full-duplex scheme always outperforms the corresponding half-duplex design for both additive and multiplicative user interferences. 
\begin{table}[h]
	\caption{SDR (in dB) performance results of the full-duplex compared to the corresponding half-duplex design for a BA-TWC with additive noise. OPTA values are also included.}
	\label{table:SDR_ANTWC_p}
	\setlength\tabcolsep{2pt}
    \def\arraystretch{1.0}
    \centering
	\begin{tabular}{ c | c c c c c c}
		\hline
		Correlation && &$\epsilon_1 = 0$ &$\epsilon_1 = 0.005$& $\epsilon_1 = 0.01$ &  $\epsilon_1 = 0.05$\\
		coefficient&r & & $\epsilon_2 = 0$ &$\epsilon_2 = 0.01$& $\epsilon_2 = 0.05$& $\epsilon_2 = 0.10$\\ \hline\hline
		$\rho = 0$&1 &half-duplex & 4.40 & 4.17 & 3.60 & 2.69 \\
		&2 &full-duplex & 9.31 & 8.18 & 6.33 & 4.34 \\
		&2 &OPTA& 12.04 & 11.27 & 9.65 & 7.35 \\  
		\cline{2-7}
		&2  &half-duplex & 9.31 & 8.18 & 6.33 & 4.34 \\
		&4  &full-duplex & 20.24 & 13.18 & 9.57 &6.47  \\
		&4 & OPTA & 24.08 & 22.54 & 18.99 & 14.45 \\ 
		\hline
		$\rho = 0.9$ &1  &half-duplex & 8.80 & 8.67 & 8.44 & 8.09 \\
		&2 &full-duplex& 11.63 & 11.06& 10.25 & 9.26 \\
		&2 & OPTA & 19.25 & 18.48 & 16.86 & 14.56 \\
		\cline{2 - 7}
		&2 &half-duplex & 11.63 & 11.06 & 10.25 & 9.26 \\
		&4 &full-duplex & 20.68 &16.76 &14.03  &11.65 \\
		&4 & OPTA& 31.29 & 29.75 & 26.20 & 21.66 \\ 
		\hline
	\end{tabular}
\end{table}

\begin{table}[h]
	\caption{SDR (in dB) performance results of the full-duplex compared to the corresponding half-duplex design for a BM-TWC with additive noise. OPTA upper bounds values are also included.}
	\setlength\tabcolsep{2pt}
	\def\arraystretch{1.0}
	\centering
	\begin{tabular}{ c | c c c c c c}
		\hline
		Correlation&& &$\epsilon_1 = 0$ &$\epsilon_1 = 0.005$& $\epsilon_1 = 0.01$ &  $\epsilon_1 = 0.05$\\
		coefficient&r & & $\epsilon_2 = 0$ &$\epsilon_2 = 0.01$& $\epsilon_2 = 0.05$& $\epsilon_2 = 0.10$\\ \hline\hline
		$\rho = 0$&1  &half-duplex &4.40  &4.17  &3.60  &2.69  \\
		&2 &full-duplex & 4.40 & 4.18 & 3.60 & 2.69 \\
		& 2& OPTA upper bound&8.35 & 7.81 & 6.71 & 5.10 \\
		\cline{2-7}
		&2  &half-duplex & 9.31 &8.18 &6.33  &4.34 \\
		&4  &full-duplex &9.31 &8.19  & 6.34  &4.34 \\
		& 4 & OPTA upper bound& 16.71 & 15.61 & 13.37 & 10.13 \\
		\hline
		$\rho = 0.9$ &1  &half-duplex  & 8.80 & 8.67 & 8.44 & 8.09 \\
		&2 &full-duplex& 9.75 & 9.52 & 9.30 & 8.81  \\
		& 2 & OPTA upper bound& 15.55 & 15.01 & 13.93 & 12.31\\  
		\cline{2-7}
		&2  &half-duplex & 11.63 & 11.06 & 10.24 & 9.26  \\
		&4  &full-duplex &12.30 &12.20 &11.07  &10.10 \\
		& 4 & OPTA upper bound& 23.90 & 22.83 & 20.59 & 17.34 \\
		\hline
	\end{tabular}
	\label{table:SDR_BMC_p}
\end{table}

For the BA-TWC with additive noise, due to the channel structure, it is feasible to perfectly ``cancel'' channel interference by subtracting the locally transmitted symbol from the channel outputs. However, in our decoding scheme we do not attempt to explicitly cancel the effect of self-interference from the channel outputs since one cannot undo the effects of self-interference in all TWCs such as the BM-TWC. Rather, we use the known self-interference, as shown in~\eqref{eq:decoders}, to condition the decoding of channel outputs. 

\begin{figure}
	\centering
	\subfigure[]
	{
		\includegraphics[scale=.5]{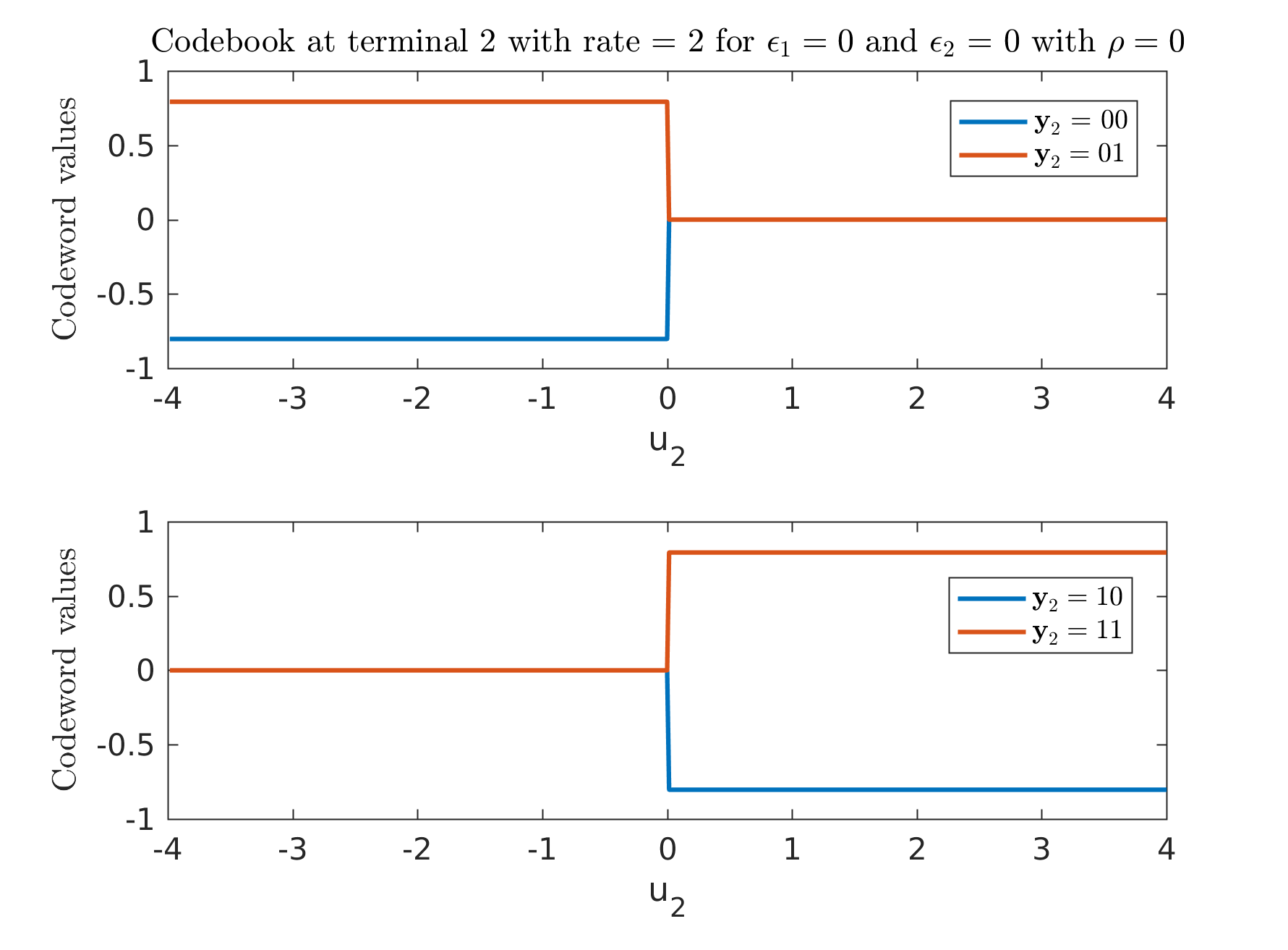}
		\label{fig:a}
	}

	\centering
	\subfigure[]
	{
		\includegraphics[scale=.5]{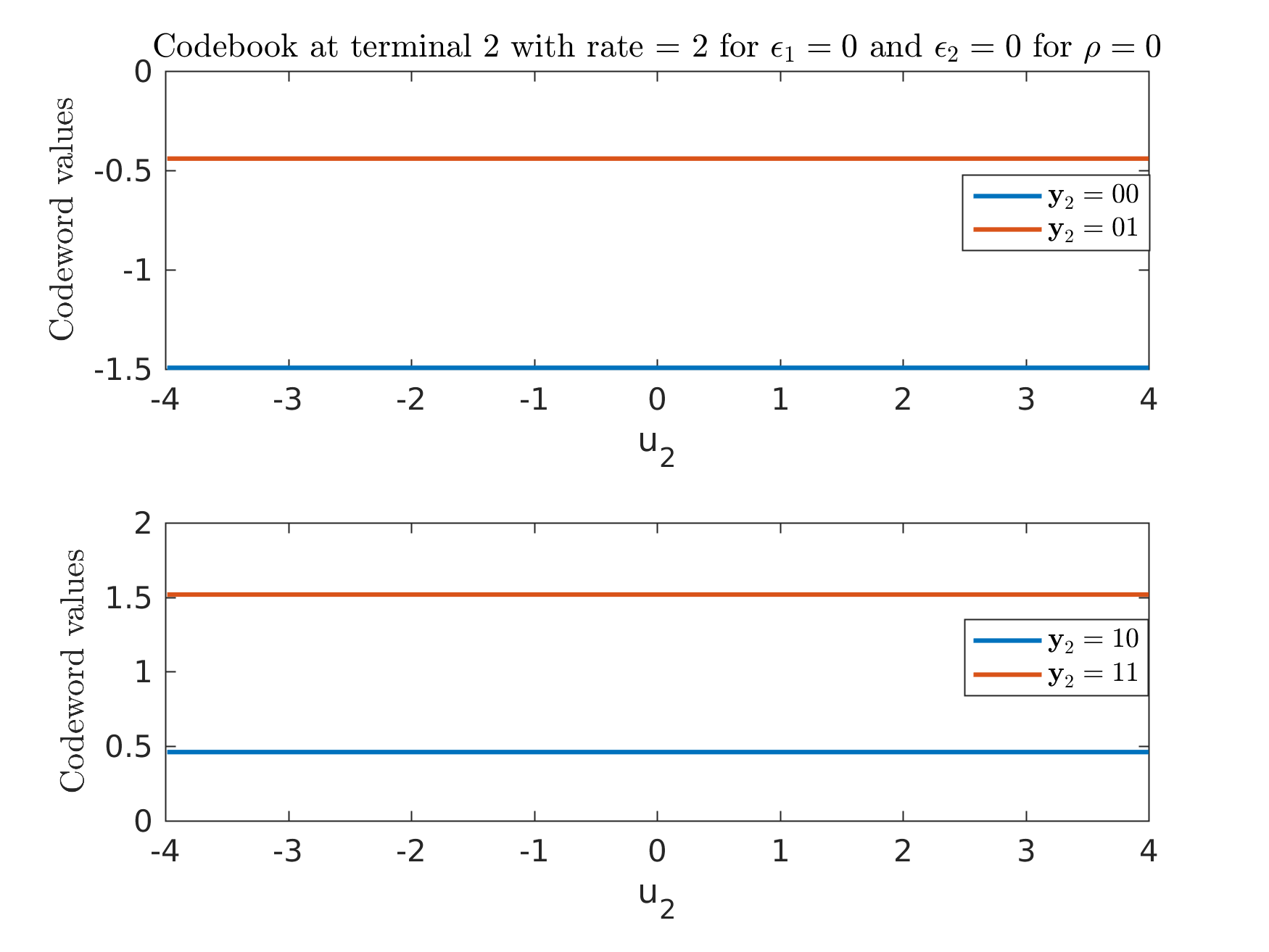}
		\label{fig:b}
	}

	\caption{Codebook constellations at terminal two for full-duplex (Fig.~\ref{fig:a}) and half-duplex (Fig.~\ref{fig:b}) schemes over a noiseless BM-TWC with uncorrelated sources ($\rho = 0$) and $2$-bit quantizers.}
	\label{fig:codebooks_H_F}
\end{figure}

Considering \eqref{eq:decoder_1}, for instance at terminal two, one can observe that the side-information $U_2=u_2$ not only does provide prior information $f_{U_1|U_2=u_2}$ for estimating $U_1$ but also determines the right channel state for decoding in the presence of self-interference, e.g., $P_{\textbf{Y}_2|\textbf{X}_1, \textbf{X}_2}(\textbf{y}_2 | \textbf{x}_1 , \alpha_2(u_2))$. For a noiseless BM-TWC, the codebook constellations of full-duplex and its corresponding half-duplex design at terminal two with uncorrelated sources ($\rho = 0$) and quantization rate $r = 2$ (bits/source sample) are depicted in Fig.~\ref{fig:codebooks_H_F}. The abrupt changes in the codebook values for the full-duplex design (Fig.~\ref{fig:a}) illustrate how the decoder tries to mitigate interference; however, for the corresponding half-duplex setup  (Fig.~\ref{fig:b}), where there is no interference as the local encoder transmits an all-one $2$-tuple over the TWC, the codeword values are constant over the entire support of the locally observed source $U_2$.   

\begin{figure}
	\centering
	\subfigure[]
	{
		\includegraphics[scale= 0.5]{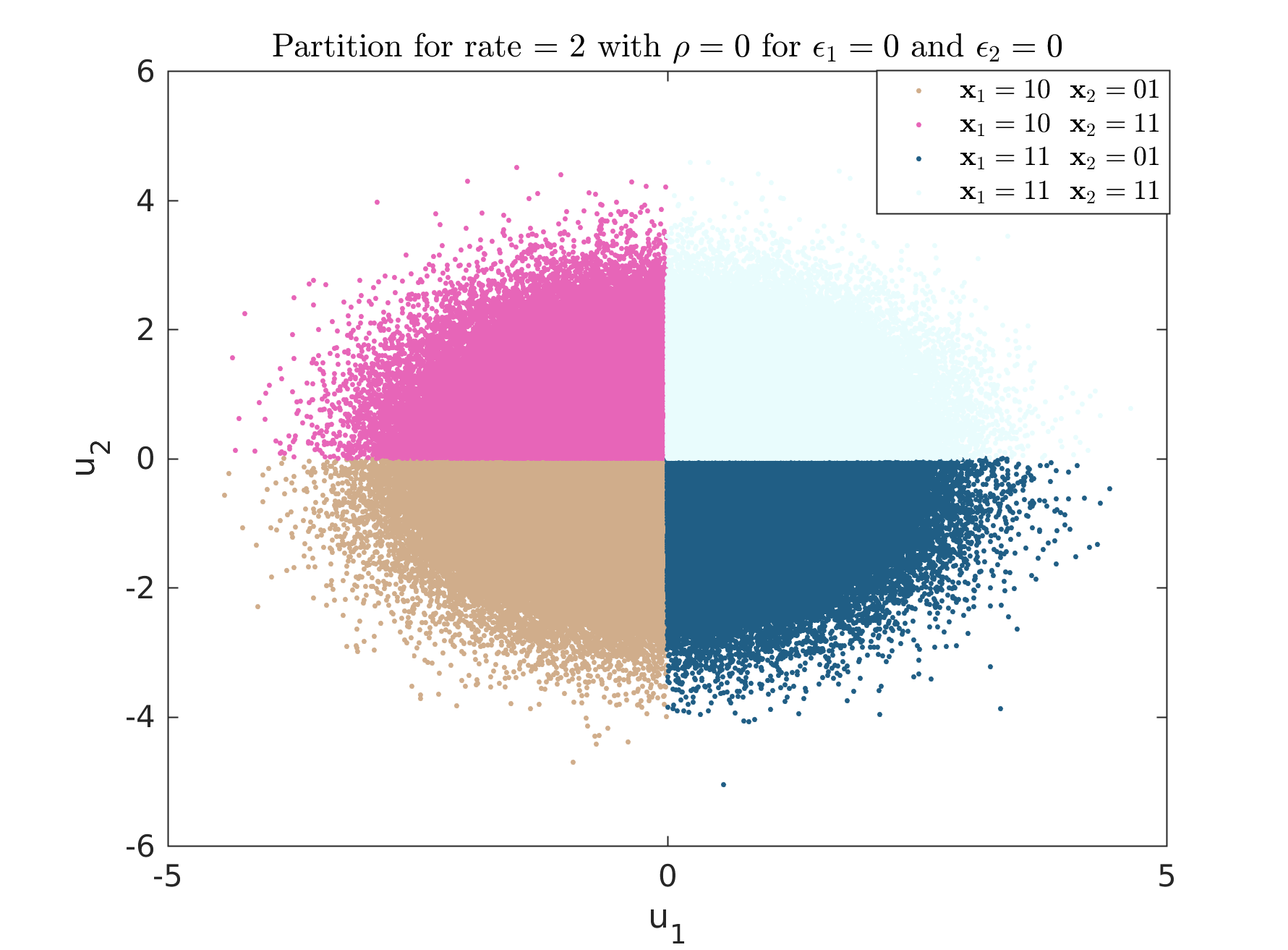}
		\label{fig:P_a}
	}

	\centering
	\subfigure[]
	{
		\includegraphics[scale=.5]{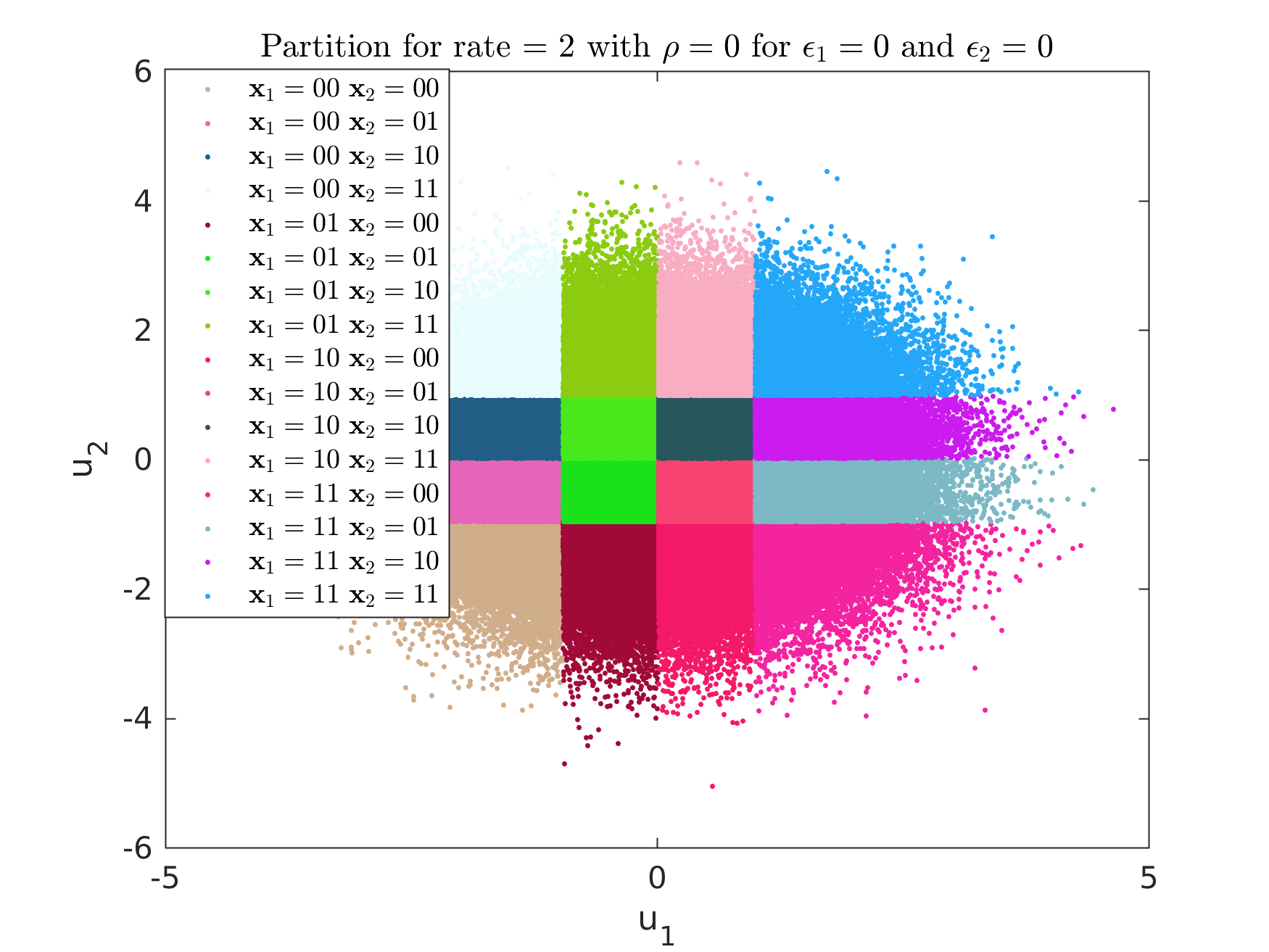}
		\label{fig:P_b}
	}
	
	\caption{Quantization cells for full-duplex (Fig.~\ref{fig:P_a}) and half-duplex (Fig.~\ref{fig:P_b}) schemes over a noiseless BM-TWC with uncorrelated sources ($\rho = 0$) and $2$-bit quantizers.}
	\label{fig:partitions_H_F}
\end{figure}

Moreover, the encoding criterion described in \eqref{eq:encoder_1}, specifies a set $S_{\textbf{x}_1}$ on the real line whose members, if mapped to the $\textbf{x}_1^{th}$  channel input compared to any other channel inputs $\hat{\textbf{x}}_1 \in \mathcal {X} ^ N$ at terminal one,  will result in lower MSE. This criterion attempts to form $\mathcal {P}_1$ as a partition of $\mathbb{R}$ such that the self-interference is avoided by taking into account all possible interference induced by user two, i.e., by averaging over all possible values of $\textbf{x}_2 \in \mathcal {X} ^ N$ in addition to combating channel noise. The encoding \eqref{eq:encoder_2} and decoding \eqref{eq:decoder_2} functions attempt to mitigate self-interference in a similar manner. For a noiseless BM-TWC, the quantization cells of the full-duplex design as well as its corresponding half-duplex scheme with uncorrelated sources ($\rho = 0$) and quantization rate $r = 2$ (bits/source sample) are shown in Fig.~ \ref{fig:partitions_H_F}. To combat interference, the full-duplex design (Fig.~\ref{fig:P_a}) does not use all partition indices, whereas, in the half-duplex setup (Fig.~\ref{fig:P_b}), where interference is perfectly avoided, the encoders can use all possible quantization indices for transmission. Finally, it is direct to note that when the sources are correlated, the SDR performance is significantly improved over the case of uncorrelated sources since the statistical source correlation boosts the decoders' reconstruction reliability.

\section{Conclusion and Future Work}\label{sec:conclusion_future}
We extended the optimality conditions in \cite{b7} for the COSQ design to the TWC setup such that the statistical dependency between the two users can be exploited as side information at the decoders. We compared the performance of our designs with half-duplex schemes that avoid interference. We showed that our proposed full-duplex system can considerably outperform the corresponding half-duplex scheme (with identical overall transmission rate) for the BA-TWC with additive noise. Also, for the BM-TWC with additive noise where the interference cannot be perfectly eliminated, the full-duplex design provides superior performance compared to the half-duplex system. Inspired by the proposed JSCC scheme in \cite{b21}, where the feedback information is used in the design of the quantizer, future work includes proposing an interactive scheme where the channel inputs, at each time instant, are chosen depending on both the source samples as well as the previously received channel outputs.   
\section{Acknowledgment}
The first author wishes to thank Jian-Jia Weng for numerous invaluable discussions on two-way communication systems. 
\bibliographystyle{IEEEtran}
\bibliography{bibFile}
\end{document}